\documentclass[letterpaper,twocolumn,10pt]{article}
\usepackage{usenix}

\usepackage{tikz}
\usepackage{amsmath}

\usepackage{tikz}
\usepackage{enumerate}
\usepackage{algorithm}
\usepackage[noend]{algpseudocode}
\usepackage{amsfonts}
\usepackage{amsmath}
\usepackage{amssymb}
\usepackage{amsthm}
\usepackage{array}
\usepackage{bm}
\usepackage{cite}
\usepackage{color}
\usepackage{dblfloatfix}
\usepackage{enumitem}
\usepackage{hyperref}
\usepackage{mathtools}
\usepackage{listings}
\usepackage{subcaption}
\usepackage{url}
\usepackage{xspace}
\usepackage[font={footnotesize},labelfont=bf]{caption}
\usepackage{flushend}
\usepackage[symbol]{footmisc}
\usepackage{balance}

\usepackage{comment}
\usepackage{booktabs}
\usepackage{svg}

\usepackage{float} 

\usepackage{enumitem}

\usepackage{macro}
\author{
    {\rm Alex Kantchelian}$^*$, 
    {\rm Casper Neo}$^\dagger$, 
    {\rm Ryan Stevens}$^*$,
    {\rm Hyungwon Kim}$^*$,
    {\rm Zhaohao Fu}$^*$,
    \\
    {\rm Sadegh Momeni}$^*$,
    {\rm Birkett Huber}$^*$,
    {\rm Cem Topcuoglu}$^*$,
    {\rm Senaka Buthpitiya}$^*$,
    \\
    {\rm Elie Bursztein}$^*$,
    {\rm Yanis Pavlidis}$^*$,
    {\rm Martin Cochran}$^*$,
    {\rm Massimiliano Poletto}$^*$
    \\[4pt]
    \begin{tabular}{c}
        $^*$Google, \texttt{\{akant,stevensr,clintk,zhaohaofu,samomi,} \\
\texttt{bthuber,cemtopcuoglu,senaka,elieb,ypavlidis,}\\
\texttt{martincochran\}@google.com}, 
        \texttt{max.poletto@gmail.com} \\
        $^\dagger$Palace Cybersecurity, \texttt{casper@palacecyber.com}
    \end{tabular}
}

\begin{document}

\date{}

\title{\Large \bf \fc: High-Precision Insider Threat Detection \\Using Deep Contextual Anomaly Detection}

\maketitle

\footnotetext[2]{Work done while the author was at Google.}

\begin{abstract}

Insiders with privileged access have the power to cause great harm to their organization. 
Even a single insider threat incident can be catastrophic, resulting in both financial losses and reputation damage.
These threats are some of the most difficult to detect, as attack activity is interspersed in large volumes of legitimate activity.
Although it is a serious threat, the literature is sparse aside from a few studies with various limitations, preventing their practical deployment in large scale organizations.

We present \fc: \underline{f}ast and \underline{a}ccurate \underline{c}ontextual \underline{a}nomaly \underline{de}tection, a high-precision, deep-learning system that has served as the last line of defense against insider threats at \ggl
since 2018.
\fc is an innovative self-supervised system that detects suspicious events by considering the context surrounding each event, including relevant facts about the user and resource involved.
It is built around a new multi-modal model that is trained on corporate document access, SQL table access, and HTTP/RPC request logs.
To overcome the scarcity of labeled incident data, \fc employs a novel contrastive learning strategy trained exclusively on benign activity.

\fc detects insider attackers with an extremely low false positive rate, lower than 0.01\%.
For single rogue events, such as the illegitimate access to a sensitive document, the false positive rate is as low as 0.0003\%.
To the best of our knowledge, \fc is the only proposed insider risk anomaly detection system with a false positive rate low enough for use in large corporate environments. 

\end{abstract}

\section{Introduction}


Insider threats are a fast-growing attack vector that organizations have to contend with. The number of companies experiencing between 20 and 41 insider-led incidents increased from 53\% in 2018 to 67\% in 2022~\cite{ponemon2022}. 
In those incidents, insiders use their \emph{authorized} access to cause harm to their organization, either \emph{wittingly} or \emph{unwittingly}~\cite{cisa-it}.
Larger organizations face more attempts and incidents as risks increase with a more numerous and distributed workforce that is harder to vet, a more complex infrastructure that offers more attack surface, and a more expansive user base that appeals to a larger group of bad actors~\cite{ponemon2022}.

With a workforce of over $100,000$ full-time employees
(and additional temps, vendors, and contractors), one of the largest technical infrastructures in the world, and a multi-billion user-base, \ggl is one of the most targeted global companies, including by insiders.
To protect its user data against threats ranging from nation-state infiltrated employees, malevolent employees trying to sell data on the black market, and well-intentioned employees being tricked or making mistakes; \ggl heavily invests in detecting and remediating insider threat attacks---including ongoing efforts to innovate and deploy ML-based anomaly detection systems.

Despite the importance of insider threats, there is a scarcity of published research on the subject.
This is partly due to the sensitivity of the topic and partly due to the technical difficulty of building robust detection systems.
%

Existing literature in this domain is often hindered by the use of unrealistic datasets that lack a representative volume of benign activity.
Consequently, many proposed systems report false-positive rates (FPR) that are operationally prohibitive for large-scale organizations. Notably, studies in the literature often report FPRs greater than 1\%\cite{li2023}. 
At our scale, such rates would produce too many spurious alerts to be usable. In large corporate environments, extremely low FPRs are crucial to ensure manageable alert precision. While this approach inherently leads to lower recall and higher false-negative rates (FNR), it is a necessary design trade-off. Without it, alert fatigue and processing bottlenecks would render the system unusable.

To help bridge the gap in the literature, this paper
presents \fc, a deep-learning anomaly detection system actively providing a last line of defense against
insider threats at \ggl since 2018.
\fc is uniquely capable of detecting anomalies at the single-event level, in contrast to many other anomaly detection systems that operate at the volumetric level. This allows \fc to operate and detect low-volume targeted attacks such as exfiltrating high-value documents.
Upon deployment, within its first weeks of operation, \fc uncovered multiple high-profile but previously unknown and unsuspected malicious events, including early-stage offensive security reconnaissance activity. 

Our design learns the relevant context around access events, using features that indicate what a principal works on and who typically accesses a given resource.
Intuitively, \fc learns which resources employees are likely to access and highlights deviations from the norm.

We use a Machine Learning architecture akin to two-tower recommendation systems~\cite{covington2016, ruan2023} to learn the relationship between principals and resources in our system.
This architecture allows us to embed semantically similar principal-resource pairs (benign activity) close to each other, while the non-similar pairs (anomalies) are far from each other.
\fc uses an innovative training technique that enables training even in the absence of incident data.


We evaluate \fc within \ggl by simulating a diverse set of real-world insider threat scenarios. This simulated attack activity is mixed with the extremely large volume of regular activity happening in the organization -- tens of billions of actions per year, even after deduplication. \fc can identify attackers with an extremely low false positive rate, lower than 0.01\%.
For single rogue events, such as the illegitimate access to a sensitive document, the false positive rate is as low as 0.0003\%.



In summary, we make the following contributions:

\begin{itemize}
    \item We present \fc, the first high-precision contextual anomaly detection system for insider threat detection that can find realistic red team attacks with a false positive rate as low as 0.0003\% for single access events.

    \item We propose a novel contrastive learning strategy that 
    reduces the unsupervised anomaly detection problem to a supervised one. Our technique, \emph{positive sampling}, 
    allows \fc to be trained using only benign data.

    \item \fc is a multi-modal model that can detect anomalous access across diverse resource types, including documents, data-stores, and internal websites.
    
    \item \fc is robust against distribution shift and generalizes to unseen-at-training-time users and resources, eliminating the need for frequent model retraining.
    
    \item \fc is a multi-scale detector. In addition to its single-event detection capabilities, it can catch red-team \textit{attackers} with a daily false positive per user budget lower than $0.01\%$.


\end{itemize}

\section{Background}

\subsection{Rule-Based Detection}

Rule-based systems are a mainstay for security teams. They enable the security engineers to express complex detection logic over the raw features of the
application domain. For instance, the Zeek (formerly Bro) network monitoring system~\cite{paxson1999} supports stateful intrusion detection, and Yara~\cite{yara-online} enables threat analysts to author static detection rules to hunt for known Tactics, Techniques, and Procedures (TTPs).

The explicit rules, when tuned to target specific known malicious behaviors, offer excellent precision and recall. However, adversary TTPs, company internal systems, and access controls are all fast-evolving, which makes writing up-to-date detection rules that cover the entire attack surface near-impossible~\cite{ali2025deep}.

This is why, in conjunction with a rule-based system, anomaly detection is needed to provide defense-in-depth and limit blind spots.
The development of such a system for detecting insider attacks has specific requirements compared to traditional anomaly detection systems.


\subsection{Enterprise Level Insider Threat Detection}

Insider threat detection is a fundamentally out-of-distribution problem.
We expect the next attack to involve different actors, targets, strategies, and overall goals from the known previous attacks.
Therefore, insider threat detection \emph{requires catching rare, novel behaviors.}

In this setting, a straightforward supervised learning approach does not succeed at recognizing novelty~\cite{hastie2001,bishop2006}.
This is because supervised learning can only detect patterns that have previously manifested in the training set -- a limitation it shares with a rule-based detection systems.
The domains where supervised learning is successfully employed are also considerably more predictable: detection of deceptive reviews~\cite{mukherjee2012}, unwanted behavior on social networks~\cite{stein2011}, and adversarial advertisements~\cite{sculley2011}. 


Traditional anomaly detection often relies on volumetric signals.
For instance, an IP hitting significantly more machines or ports than normal can be a sign of network reconnaissance activity.
This is unsuitable for our application domain as it will only catch egregious attacks after the fact, such as the exfiltration of a large volume of documents.
Such an approach fails to detect low-volume targeted attacks aimed at exfiltrating a single critical piece of data, such as chip design plans, financial information, or proprietary technical designs.
Hence, enterprise level insider threat detection requires \emph{detection granularity down to single resource access.}

Lastly, traditional anomaly detection systems infamously suffer from high false alarm rates and security-irrelevant alerts~\cite{sommer2010}.
Since the produced alerts are ultimately reviewed by security analysts who are expensive to hire and train, the detection system must operate within a stringent alert budget.
Unfortunately, even at seemingly low false positive rates, say 0.1\%, the large base rate of events in the activity logs of a large-scale company, say 10 million events per day, will result in 10 thousand alerts per day.
This is an unmanageable volume for a typical size security team and will quickly result in analysts ignoring the alerts.
At scale, the vast amount of legitimate activity together with the rarity of insider attacks \emph{require extremely low false positive detection rates.}

\subsection{Research Gaps}
Prior work on insider threat detection suffers from data sets with unrealistic benign activity, false positive rates that are unsuitable for large enterprises.

To the best of our knowledge, no publicly available data set is suitable for developing insider threat detection systems in large-scale organizations.
The CERT Insider Threat Test Dataset~\cite{glasser2013,cert-data} is one of the most pervasively used data sets.
It is created by randomly sampling benign user activity under probabilistic assumptions.
For any real large-scale enterprise, however, we expect and in practice observe, that the benign behaviors possess a diversity and unpredictability that would be extremely challenging to replicate under the random sampling schemes used for the CERT data set.
Other works completely eschew the question of generating realistic benign activity and focus instead on creating richer, more representative attack instances~\cite{yuan2020, pereira2023}.

Unfortunately, a significant number of insider threat works rely on the CERT data set~\cite{yuan2018,randive2023,li2023,pal2023}.
There, the highly predictable benign activity may result in under-estimated false positive rates in practical large-scale organization deployments.
Even without this issue, the reported false positive rates are often on the order of 1\%, which would be unacceptable in our environment.

In summary, neither does an appropriately realistic insider threat data set exist for comparison, nor are traditional anomaly detection systems applicable in our environment.

\subsection{Self-Supervised Learning}


In this section, we give a brief and non-security specific overview of the relevant machine learning techniques used by \fc.
These techniques fall under the umbrella of \emph{self-supervised learning}~\cite{agrawal2015,doersch2015}, which forgoes the need for labeled and representative attack training data, while making use of the entire set of available but unlabeled data.

Contrastive learning~\cite{hadsell2006} is one of the most effective approaches to self-supervised learning.
It proceeds by synthetically generating (sampling) a new class of training examples from the existing data points,
and thereby reduces an apriori unsupervised learning problem into a supervised one.
A supervised classifier can then be trained to distinguish between the two classes, and possibly learn some intermediary representations for the data points, a.k.a. embeddings, in the process.

The two-tower recommendation system~\cite{covington2016, ruan2023} is a staple of the item retrieval domain.
Each tower maps its input into a fixed-size high dimensional vector, a.k.a. an \emph{embedding}.
One tower would typically process a query (e.g. the raw text of the user's search query), the other an item (e.g. a video).
For a given notion of distance in the embedding space, these two-tower models are trained such that the embeddings of ``matching'' query-item pairs are close to each other, while ``mismatching'' pairs are far from each other.

\emph{Negative sampling}~\cite{mikolov2013} is a specific form of contrastive learning that is widely used for training large-scale modern recommendation systems.
Negative sampling requires a data set of matching query-item pairs (the positive class) and forms synthetic pairs (the negative class) by randomly sampling the elements of the pair.
The two-tower model weights are then adjusted so as to push the matching embedding pairs together, and the mismatching embedding pairs away from each other.

\section{Research Statement}

Operating at a scale of 100,000 employees and billions of daily events, an organization like \ggl requires a solution that scales not only in data throughput but also in operational viability.
At this magnitude, even a false-positive rate of 0.1\% would generate millions of daily alerts, effectively burying security analysts in noise.
For an anomaly detection system to be useful in such an environment, we need a detection system with low false-positive rates that detects both low-volume targeted attacks and novel threats, in the absence of malicious data for training.

We first approach the problem from the security analyst's perspective.
During an investigation, security analysts need to collect all relevant facts surrounding an event.
For example, a user viewing a sensitive file can only be properly understood by knowing the the user's role, team, and projects.
If the user is viewing a file owned by a team they are collaborating with, the event could be deemed normal; however, if there is no collaboration history between the user or the user's own team, and the target team, the event might be deemed anomalous, justifying further investigation.

This approach is, intuitively, a contextual anomaly detection process: the analyst checks the validity of the event by looking at the context around it.
We hypothesize that if we can replicate and scale up the work of the analyst by limiting the context to a relevant and appropriate subset of the state of the world, we will have developed an effective methodology.

To develop this system, we assume that the overwhelming majority of events within the organization are benign.
This allows us to effectively apply self-supervised learning trick by creating synthetic, non-existent patterns that represent anomalies, using only the available unlabeled but assumed-benign data.

Our \textbf{threat model} for this insider detection system consists of two attack scenarios:

\begin{enumerate}
\item A rogue employee abuses their authorized access to critical resources.
\item A non-malicious employee is misled or tricked into accessing critical resources (e.g., when an external party compromises employee credentials and uses them to perform malicious actions).
\end{enumerate}

In this threat model, the rogue or compromised accounts display behaviors that are atypical for an employee with the same role and access.
For example, such a compromised account may search corporate documents for valuable digital assets such as development plans or trade secrets~\cite{chen2014}.

To ensure the practical applicability of such a system in an enterprise environment, we explore the following research question:
\textbf{Is it possible to develop an insider detection system that detects insider threats across multiple scales, from single-event to days of activity, within an operational budget of ten audits per day, while only using benign data for training?}




In Section~\ref{sec:design}, we present the design of \fc, where we make the assumption that the problem at hand is a supervised learning task.
Later, in Section~\ref{sec:cont-learn} we clarify this point, and explain how we train the system in the absence of labeled attack data, by reducing the unsupervised problem into a supervised learning task.
In Section~\ref{sec:imp}, we provide details of our implemented architecture and featurization.

\section{\fc Design}
\label{sec:design}

\fc finds anomalies by looking for events that are statistically unlikely given their context. For each event, it produces a score which can rank how unusual that event is.

Our system ingests raw activity logs, and creates featurized events which are then used to train the model.
From the featurized events, the model generates anomaly scores.
These scores are used in the detection of insider threats.

For the rest of the paper, we use the following definitions:

\textbf{\textit{Principals.}} The set of all unique human identities in the system, including full-time employees, temporary workers, vendors, and contractors.

\textbf{\textit{Resources.}} The set of all unique entities accessible by \emph{principals} in the system, such as documents, SQL tables, and network endpoints.

\textbf{\textit{Event.}} A \emph{principal} interacting with a \emph{resource} at a specific point in time. The environment consists of a continuous stream of \emph{events}.

\textbf{\textit{Action.}} The facts available to featurize a \emph{resource} at the \emph{event} time.
The action is computed using the information about the resource, specifically the set of principals who have previously accessed the resource. Section~\ref{sec:historical-featurization} presents the featurization scheme in detail.

\textbf{\textit{Context.}} The facts available to featurize a \emph{principal} at event time.
The context is computed using the information about the principal's profile (e.g., role, team, project assignment) and social interactions (e.g., meetings). The details are provided in Section~\ref{subsec:context-types}.

\fc produces anomaly scores for events, which are pairs formed by a context and an action.
Let $\bA$ be the set of all actions and $\bC$ be the set of all contexts.
Our goal is to train a machine learning system $f:\bA\times\bC\to\bR$, which maps contexts and actions to anomaly scores.
We adapt the two-tower model architecture~\cite{covington2016} and decompose $f$ into a \emph{context} tower, $E_C:\bC \to \bR^d$,
and several \emph{action} towers, $E_A^t:\bA \to \bR^d$, one for each action type, $t$. 
While each action tower is a separate deep neural network, for brevity,
we use $E_A$ to refer to them collectively,
applying the tower of the appropriate action type.
These towers map contexts and actions to \emph{context embeddings} and \emph{action embeddings}.

\begin{figure}[tb]
\centering
  \includegraphics[width=1.0 \columnwidth]{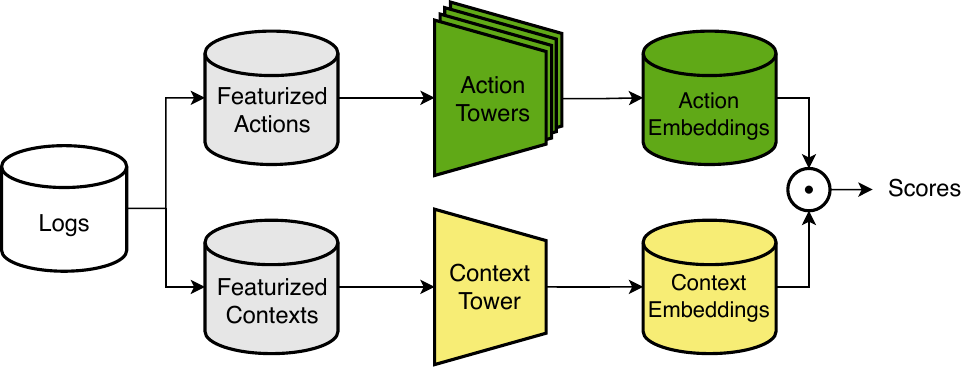}
\caption{
\small
The core idea behind \fc. Events are split into contexts and actions, the model computes embeddings independently for each, and the final anomaly score is obtained as the cosine distance, essentially a dot product, between the context and action embeddings.
}
\label{fig:overview}
\end{figure}

We train the model to embed contexts and actions close to each other if and only if they correspond to a benign event, and define anomaly scores to be the cosine distance score between the embedded action and context, $f(a,c) = d_{cos}(E_A(a), E_C(c))$, so benign and suspicious events correspond to low and high scores respectively. The precise training procedure is explained in Section~\ref{sec:cont-learn}.
%
%
As a natural consequence of this training, the embeddings capture behavioral semantics: pairs of principals who access similar resources, and pairs of resources accessed by similar sets of principals, will have a with a low cosine distance between their embeddings.
Figure~\ref{fig:overview} summarizes the core idea behind \fc.
%





$f$ can only produce a single score out of each context and action pair.
This is useful when focusing on a specific principal and their most suspicious events, or for direct, single-event detection purposes.
This detection mode is appropriate for fast detection of egregious-enough single accesses.
However, because most attacks consist of multiple suspicious but not necessarily egregious events, attacks can become easier to detect when considering many events together.
We thus extend \fc's core single-event scoring capability to score an arbitrary set of events, such as all actions of a principal for a specific time period (e.g., a week).

We introduce two techniques that take advantage of our embedding space, where cosine distance can be used to measure behavior similarity.
First, we present \emph{common event filtering} where we filter behaviorally similar events that were performed by distinct principals, as those are heuristically benign. Then, we discuss \emph{per-principal aggregation} which groups events related to the same potential attacker -- highlighting sets of actions containing behaviorally diverse anomalies. This not only reduces the number of tickets that need to be looked at but also collocates activity from a single principal when reviewed by an analyst. Taken together, these techniques reduce system level false positive rates. 

\subsection{Common Event Filtering}
\label{sec:common-filtering}
When distinct principals with similar coworkers access similar resources at similar times,
there is often a common and justified business purpose that caused those events.
We call such events \emph{common events}.
We expect $f$ to learn to produce low scores for such events.
However, even rare inference errors are visible in the tail at scale.

Since we are interested in malicious activity,
we remove high scoring common events from the detection set.
This is safe from adversarial manipulation,
assuming it is significantly more difficult to compromise multiple principals than to compromise just one.
Common event filtering shares this assumption with multi-party authorization systems.

We measure the similarity of coworkers and resources using the embedding distance of
context and action embeddings respectively.
Formally, given thresholds for similarity $T_a, T_c\in (0,1)$ and multiplicity $m\in\mathbb N$,
we remove an event $(a_0, c_0, p_0)$ if and only if
there exists events $(a_1, c_1, p_1), ..., (a_m, c_m, p_m)$
such that the principals are distinct,
\[
    \forall p_i, p_j \in \{p_0..., p_m\},
    \hspace{3mm}
    p_i \not= p_j
\]
the contexts are behaviorally similar,
\[
    \forall c_j \in \{c_0..., c_m\},
    \hspace{3mm}
    d_{cos}(E_c(c_0), E_c(c_j)) < T_c
\]
and the actions are behaviorally similar,
\[
    \forall a_j \in \{a_0,..., a_m\},
    \hspace{3mm}
    d_{cos}(E_a(a_0), E_a(a_j)) < T_a 
\]


\subsection{Per-principal Aggregation}
\label{sec:agg}
Single action-context pairs may not be compelling enough to warrant further investigation, and
evaluating large sets of related actions amortizes the cost of understanding the principal and resources involved.
Hence, we develop the aggregator, $g: \wp(\mathbb{A}) \times \mathbb{C} \rightarrow \mathbb{R}$, 
where $\wp(\cdot)$ is the power set function,
which
scores arbitrary sets of actions performed by a single principal
and facilitates comparisons across principals.
An investigator may choose a family of actions to investigate, group the actions by principal,
produce an aggregated score for each principal and group of actions, then use the scores to inform further investigation.

We now state three requirements for any such aggregator:


\noindent \textbf{Monotonicity.}
We require that for a given context, no amount of benign activity can mask the presence of suspicious actions.
We therefore introduce a monotonicity requirement on $g$.
Specifically, for any fixed context $c$, we require the function
$A \mapsto g(A, c)$ to be non-decreasing with respect to the set
inclusion relation.
Formally:
\[
\forall c \in\mathbb{C}, \forall A', A \in \wp(\mathbb{A}),\; A' \subset A \Rightarrow g(A', c) \leq g(A, c)
\]
This constraint ensures that an attacker cannot manipulate the aggregated score downward by introducing additional actions.

\noindent \textbf{Consistency with Single-Action Scoring.}
For fixed $A\in\wp(\mathbb{A})$ and $c\in\mathbb{C}$, we require the function $a \mapsto g(A \cup \{a\}, c)$
to follow the score ordering induced by $f$.
Formally:
\begin{align*}
\forall a, a' \in \mathbb{A},\; &f(a,c) \leq f(a',c) \\
  &\Rightarrow g(A \cup \{a\}, c) \leq g(A \cup \{a'\} , c)
\end{align*}

\noindent \textbf{Approximate Idempotence to Behaviorally Redundant Actions.}
\label{sec:idempotence}
With the above requirements alone, if $f$ always produces non-negative scores, $g$ may be defined to be the sum of individual action-context scores:
$g(A,c) = \sum_{a\in A}f(a,c)$.
This scoring method is undesirably biased: large action sets $|A|$ will tend to score higher than small ones.

Low-level logs often contain large amounts of behavioral redundancy.
For example, loading a single URL can result in a large number of subsequent implied requests.
A detection system biased by $|A|$ would have an unacceptably high FPR.
Therefore, we require $g$ to discount redundant actions and highlight
$A$ when it is a diverse set of anomalous behaviors.
Formally, there should exist small constants $\delta, \delta' > 0$ such that:
\begin{align*}
    \forall A\subset \mathbb A,
    & \forall a' \in \mathbb ,
    \forall c\in\mathbb C,
    \\
     & \min_{a\in A}d_{cos}((E_A(a), E_A(a')) < \delta
    \\
    &\Rightarrow g(A \cup \{a'\}, c) - g(A, c) \leq \delta'
\end{align*}


\begin{figure*}[t]
  \includegraphics[width=1\textwidth]{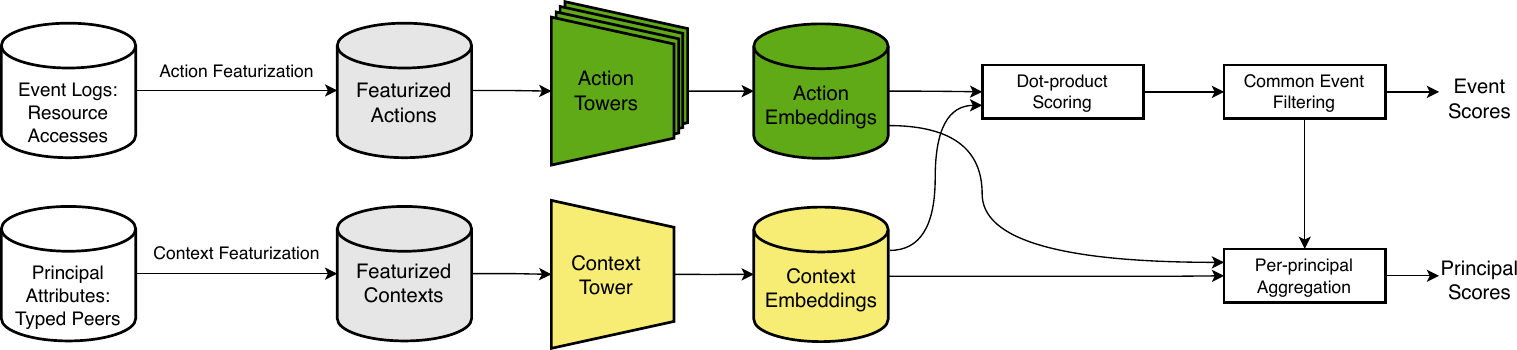}
  \caption{\fc system at inference time. Cylinders are data sets, and trapezoids are (parts of) learned ML models.}
  \label{fig:system-diagram}
\end{figure*}

\noindent \textbf{A Clustering-Based Implementation.}
\label{clustering}
Our scoring function uses hierarchical agglomerative clustering~\cite{monath2021}.
We use cosine similarity over the single
action embeddings, cosine distance between clusters centroids and specify fixed numbers for the maximum number of clusters and the maximal admissible distance for merging two clusters.
Clustering organizes $A$ into set of actions $X \subset G$, and we define the score as the sum of the max scores for each cluster:
\(
    g(A, c) = \sum_{X\in G}\max_{a\in X} f(a, c)
\)

Note that this technique does not perfectly adhere to the requirements above.
While it is consistent with single-action scoring, one can construct counter-examples to the monotonicity and approximate idempotence property.

Despite its simplicity and imperfect adherence to the above principles, we find that the clustering approach is effective at detecting insider threats.

When presenting results to security analysts, the clusters are also
useful for organizing large sets of actions by behavioral similarity.

\subsection{System Summary}
We now completed the details of our system for single-event and multi-event levels. Figure~\ref{fig:system-diagram} depicts \fc in the inference time with full details. Event logs are featurized into context and actions. Then, we embed them with context and action towers, yielding context embeddings and actions embeddings. We score the event by taking the embeddings' dot-product. Finally, we apply common event filtering and apply per-principal aggregation to create per-principal scores.


\section{\fc Training}
\label{sec:cont-learn}


In our environment, the overwhelming majority of events are benign.
While it is possible to identify subsets of events that 
are considered suspicious or outright malicious, this subset is negligible in volume,
and may be filtered when identified.

In what follows, our training data set is comprised only of
unlabeled historical events, assumed to be benign.
Despite the apparent unsupervised nature of the learning problem,
we can cast it into a supervised one using contrastive learning.

\begin{figure}[t]
  \includegraphics[width=\columnwidth]{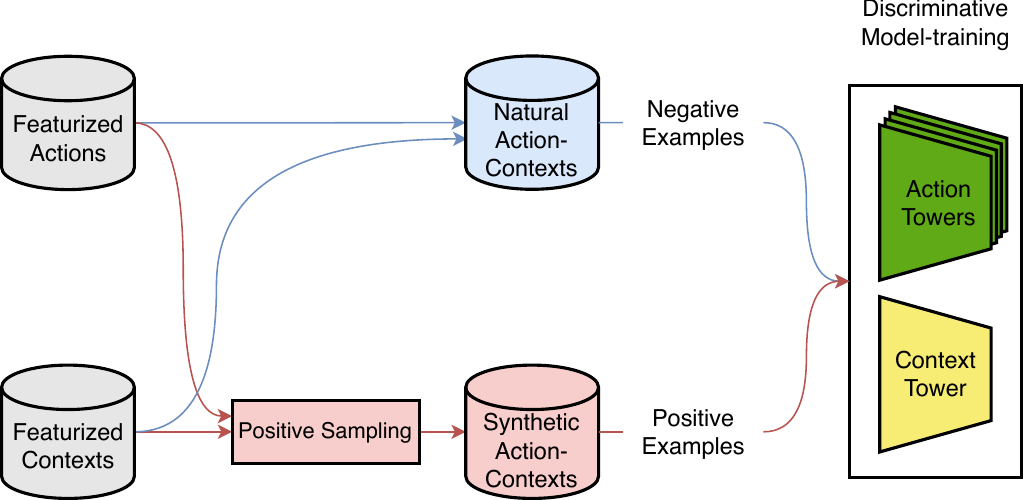}
  \caption{\fc system at training time. \fc employs a self-supervised, contrastive training strategy: the model is optimized to differentiate between natural and positive sampled pairs of action-contexts.}
  \label{fig:system-training-diagram}
\end{figure}

%
Figure~\ref{fig:system-training-diagram} presents the \fc training process.
%
We now explain the details of, and theory behind, positive sampling.

\subsection{Positive Sampling}
\label{subsec:pos-sampling}

Let $\mathcal{D} = \{(a_i, c_i, p_i)\}_{0 \leq i < n}$ be the set of all available historical
events. Each triple $i$ is a single action
$a_i\in\bA$, from a principal $p_i$, with their context $c_i\in\bC$.
To train our classifier, $f:\bA\times\bC\to\bR$, in a supervised way, we borrow an idea from
Natural Language Processing (NLP) known as negative sampling~\cite{mikolov2013}.
Negative sampling in NLP aims to find the most
similar pairs of elements.
By contrast, our approach which we call \emph{positive sampling} aims to find the most surprising, or dissimilar action-context pairs.

We create new action-context pairs by randomly mismatching actions
and contexts belonging to different principals.
Our synthetically labeled data set is defined as:
\begin{align*}
\tilde{\mathcal{D}} = &\{(a_i, c_i, -1) \;|\; 0 \leq i < n \} \\
\cup &\{(a_i, c_j, +1) \;|\; 0 \leq i, j < n; p_i \neq p_j \}
\label{check-prin}\tag{$\star$}
\end{align*}

We refer to the first set in the union as the negative pairs or natural instances
and the second as the positive pairs or synthetic instances.

For example, for a toy data set containing two events
$\{(a_1, c_1, \verb|alice|), (a_2, c_2, \verb|bob|)\}$,
we would generate exactly two natural pairs $\{(a_1, c_1, -1), (a_2, c_2, -1)\}$
and two synthetic positive pairs $\{(a_1, c_2, +1), (a_2, c_1, +1)\}$.

As there are generally $O(n^2)$ potential synthetic positive pairs for
a data set of $n$ natural events, and $n$ exceeds $10^{10}$  in our
application after data set simplification,
we avoid explicitly constructing all positive pairs.
Instead, we randomly sample $n_p$ synthetic pairs per natural pair within each training minibatch.
We sample $n_p = 10$ positives per negative at training time
and $n_p = 1$ at validation time.
This is computationally very efficient because the action and context embeddings
are computed once over a batch of natural action-context pairs. The scores of
synthetic positives are computed by sampling permutations over the context or action
embeddings and computing the cosine distances for the resulting pairs.

\subsection{Positive Sampling Learns Lift}
\label{subsec:lift}

We theoretically demonstrate that learning on positive sampling results in a
semantically meaningful classifier $f$.
Ranking events with $f$ is equivalent to sorting by $\frac{p(a, c)}{p(a)p(c)}$, i.e., the ratio of the joint probability over the product of marginals.
This quantity is know as \emph{lift} in the data mining literature \cite{coppock2002lift}, and the result holds independently of model architecture, featurization, and loss function under mild conditions.

For clarity of the proof, we use the following simplifications.
First, we randomly and independently sample action and contexts from the natural pairs without 
verifying that the principals are distinct, i.e.,
without checking $p_i\neq p_j$ in \eqref{check-prin}.
We call that distribution $\tilde{\mathcal{D}}_{\textnormal{simple}}$. 
Second, while in general, $\bA$ and $\bC$ may be infinite and richly-featurized
action and contexts spaces, this part assumes they are finite
and opaque.


\begin{theorem}[positive sampling + pointwise loss $\Rightarrow$ lift]
\label{thm:lift}
Let $\bA = \{1,..., n_a\}$, $\bC=\{1,...,n_c\}$, and $P$ be probability distribution
over $\bA\times \bC$.
Let $\ell: \mathbb{R} \rightarrow \mathbb{R}$ be a decreasing, strictly convex
differentiable function where $\ell'(t)\rightarrow 0$ at infinity.
A binary classifier $f:\bA\times\bC\to\bR$ which minimizes the expected risk under $\ell$, e.g., $\mathbb{E}_{(a, c)\sim P}[\ell(f(a, c))]$) and 
simplified synthetic positives, $\tilde{\mathcal{D}}_{\textnormal{simple}}$, produces scores
such that $\forall a,a'\in\bA \forall c,c'\in\bC$:
\[
f(a, c) < f(a', c') \Leftrightarrow \frac{P(a,c)}{P(a)P(c)} > \frac{P(a',c')}{P(a')P(c')}
\]
where we used the following notational shorthands to denote the
marginals of actions and contexts: $P(a) = \sum_{c \in \mathbb{C}} P(a, c)$,
$P(c) = \sum_{a \in \mathbb{A}} P(a, c)$.
\end{theorem}
See Appendix~\ref{sub:proof} for the proof and Appendix~\ref{subsec:lossfn} for our choice of loss function.

The lift captures a notion of affinity between a context and an action that is independent of the underlying frequencies of either.
Applying the definition of conditional probability,
\(
\frac{P(a,c)}{P(a)P(c)} = \frac{P(a|c)}{P(a)}
\).
We can see that the lift is
an action's conditional probability given its context, normalized by the probability of the action.
An action being rare, in and of itself, is not a sufficient condition for
anomalousness according to a perfectly trained model.
It must be relatively rare given the context.
Analogously, the lift equals $\frac{P(c|a)}{P(c)}$ so principals with uncommon contexts
are not unfairly highlighted by the model, they have to have performed an unexpected action.
What matters is the affinity between an action and context pair:
The probability of sampling the pair from the joint distribution over sampling them independently from the marginals.


\subsection{Robustness to False Positives and Negatives}
\label{sec:robustness-claim}
\noindent \textbf{Robustness to False Positives.}
Because of the random mismatching procedure, some synthetic pairs can mimic legitimate
activity and should consequently be considered as negative pairs.
This can happen even when care is taken to only cross pairs that belong to different
principals.
For instance, we may cross two negative pairs belonging to two employees working
under the same team where members fulfill identical job roles.
If the employees are functionally indistinguishable, there is no reason to expect a
high score for the resulting synthetic pair, and we have effectively introduced a
false positive label error.
Theorem~\ref{thm:lift} shows that, asymptotically,
these labeling errors do not matter and we still learn the lift.

\noindent \textbf{Robustness to False Negatives.}
While our training strategy assumes the natural access logs are benign,
we know there will be a subset of malicious accesses.
Such events will be present in any non-trivial, sufficiently large system.
Aside from malicious intent, human and software errors, such as configuration mistakes,
induce false negatives in our training data.
We show in our evaluation that small amounts of false negatives, both random noise and
malicious activity, are empirically tolerable.

We postulate that even when subsets of known previous malicious events are available,
a direct supervised learning approach on the labeled data is suboptimal compared
to our synthetic positive approach.
Our threat model is characterized by rare and novel abusive behavior.
We do not expect the historically malicious subset, by itself, to provide enough
behavioral diversity to teach a model to generalize to future, out of distribution,
malicious events.
This argument is independent from the challenge of
learning on highly imbalanced labeled data.

Lastly, we describe the implementation of \fc and the features we use for contexts and actions.

\section{Implementation}
\label{sec:imp}
\label{sub:retraining}

The data processing pipelines are implemented in C++ Dataflow~\cite{dataflow}
and the model is implemented in TensorFlow~\cite{tensorflow2015-whitepaper}.
\fc contains on the order of 10,000 lines of code.

Altogether, creating the training set takes a couple of days and training the model from
scratch takes a few hours with one NVIDIA H100 GPU.
By contrast, inference is fast. Detection latency is dominated by log delays
and batch job scheduling.
We also provide on-demand low-latency evaluations of $f$ with slightly stale data 
for arbitrary, previously seen, resources and principals
by caching their most recent embeddings 
and computing $d_{cos}$ on the fly.

Note that \fc can work with a variety of featurizations. We present it as it is implemented at \ggl.

\subsection{Action Featurization}
\label{sec:historical-featurization}
We model every action type of interest as an access to a resource
(e.g., a document, a binary file, an HTTP request, a presentation),
and require unique resource identifiers.
If an event involves more than one resource, for instance a SQL query involving more than one
table, we create an atomic action for each distinct resources involved.
Because resources and their associated access modality (e.g., view, execute, share)
are heterogeneous, we adopt a metadata-only featurization, ignoring both content and
access modality, and obtain a general compatibility score
between a user and a resource.

\begin{figure}[tb]
\centering
  \includegraphics[width=.73 \columnwidth]{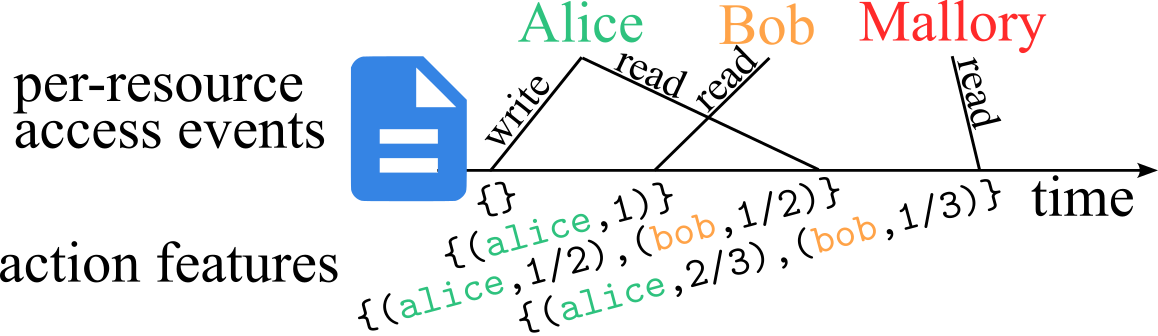}
\caption{
    \small
    History-based action featurization for a single resource.
    Timestamp and access modality are not featurized.
    Any action is represented by the weighted set of all previous accessors,
    where the weights are proportional to the frequency of accesses and sum to 1. Note that the features representing the resource changes every time a new access occurs.
}
\label{fig:history}
\end{figure}

Resources, and to some extent access modalities, form an open universe.
We cannot simply learn an embedding for each resource identifier,
as there would be no such embedding for new resources, which we expect, continuously.
Instead, given a resource and at point in time, we create an embedding
as a function of the weighted set of users who have previously accessed the resource,
as illustrated in Figure~\ref{fig:history}.

The open universe of users is considerably smaller and more stable in time than
the open universe of resources. This featurization strategy is effective at representing
resources as their access control requirements evolve, and
it gracefully handles new resources, not present at training time.

When the data is available,
\fc and history featurization is sufficiently generic that
implementing new action types can be remarkably easy.
We need only implement a parser that extracts (resource, principal, time) access triples
from some authoritative log source, update configuration, and retrain the model.
\label{subsec:action-types}
We present results for an implementation which covers the following action types:
\begin{itemize}
    \item \textbf{Cloud-based document accesses:} Documents include presentation slides,
        spreadsheets and any arbitrary files. Every such item has a unique identifier.
    
    \item \textbf{SQL-based data accesses:}
        This covers all registered data sets in a company-wide data lake.
        We ignore the SQL query statement and use the table's
        unique identifier. When there are multiple tables in the query, then
        we create one action per table.

    \item \textbf{HTTP/RPC requests to internal systems:}
        The resource identifier for HTTP requests is the hostname.
        For well-known internal sites
        we \emph{specialize} the resource identifier by parsing the URL.
        For instance, in the internal issue management service, we use the ticket
        number.
        For Remote Procedure Calls (RPCs),
        the resource identifier is the RPC service and method name pair.

\end{itemize}





\subsection{Context Featurization}
\label{subsec:context-types}


Contexts represent what is known about the user who performed the action,
at the time of the action. 
Collectively, these features enable inference of the principal's role,
in the spirit of role-based access control systems \cite{rbac1996}.
They contain a few categorical features that describe
non-sensitive, internally-public information about the user. This includes their
job family type and their discretized employment duration.
The main features are the user's weighted set of peers,
representing the implicit social networks revealed by the logs of collaboration tools. For each user, we use:

\begin{itemize}
    \item \textbf{Meeting peers.}
    The weights are proportional to the number of shared meetings and inversely proportional
    to the number of meeting participants.
    Meetings marked as private or confidential are ignored.
    \item \textbf{Code-review peers.}:
    The weights are proportional to the frequency and recency of code reviews provided or received.
    This feature describes whether a user is technical, and if so, who they work with.
    \item \textbf{Management peers.} The set of peers who share a manager or grand manager. Peers who
    only share a grand manager are assigned half the weight of those who share a manager.
    \item \textbf{Cost center peers.} Their cost center peers, weighted equally.
    Cost centers reflect high-level business areas, which often 
    go beyond grand managers and job titles.
\end{itemize}

Context features evolve slower than action features.
The largest change occurs when a user switches teams or job assignments.
Notice that the model $f$ does not rely on the identity of the principal 
associated with a context at training time.
\fc trivially handles team changes and new users without model
retraining: the model is always fed up-to-date, explicit facts about the principal's role
and coworkers.
In practice, we refresh the users' context features daily.


\subsection{Model Architecture}
\label{section:architecture}
\begin{figure}[!h]
\centering
  \includegraphics[width=.68 \columnwidth]{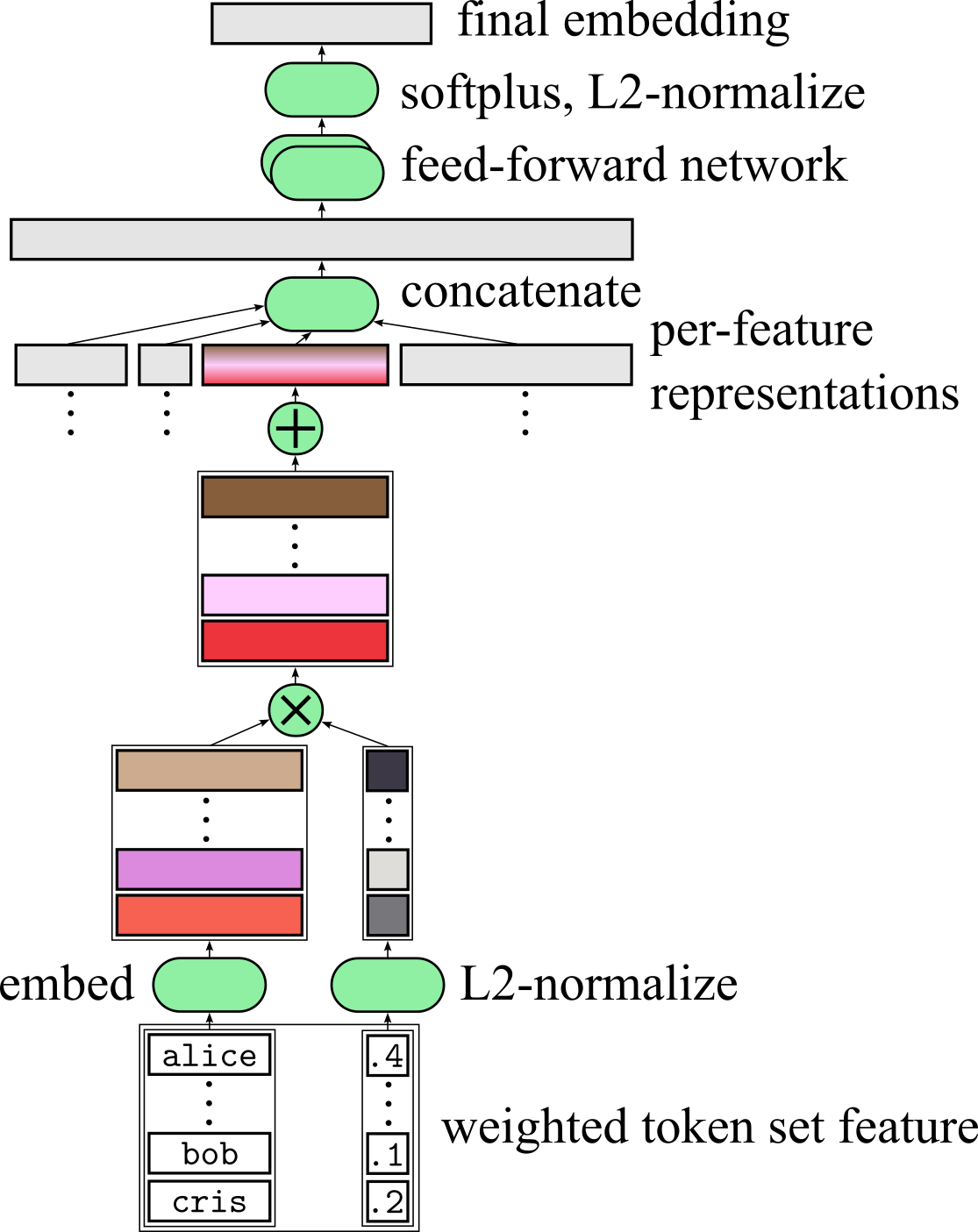}
\caption{
\small
\fc tower architecture.
Green rounded shapes denote transformations,
and rectangles denote concrete values. The subarchitecture
that reduces a variable-length weighted token set feature
to a fixed-size intermediary representation is expanded.
By convention, our final embeddings are non-negative and L2 normalized.
}
\label{fig:arch}
\end{figure}

Figure~\ref{fig:arch} presents our tower architecture.
All towers share the same architecture and input types, though their specific number of layers and neurons are allowed to vary.
Each resource type (documents, sql tables, urls) has its own \emph{action} tower.

The context and action towers all take, as input, weighted sets of employee username tokens as presented in Section~\ref{subsec:context-types}.
For actions, the tokens represent the users who recently performed the action, weighted by frequency (see Section~\ref{subsec:action-types}).

\subsection{Data Set Simplification}
\label{subsec:simplify}

Starting from the internal log sources, we separately compute the daily contexts and all actions for all users.
To accelerate the data processing pipelines and the learning process itself,
we store the context for each principal and one action per distinct accessed resource in each non-overlapping two hour period.
We join the contexts and actions in a temporally consistent manner, that is,
we pick the context with greatest timestamp earlier than the two hour period in which the actions occur.

We drop actions representing the access to company-wide
resources (those accessed by over 2000 distinct principals a day)
as they are uninteresting from a security standpoint.
Similarly, we drop actions with empty histories, which represent the first access to a resource.

Only the natural action-context features are stored on disk
because synthetic pairs are sampled from mini-batches at training time.

Lastly, we apply hyperparameter optimization using a spatially and temporally split validation set. These details are presented in appendices~\ref{sec:hparam} and ~\ref{sec:spatial-split}, respectively.

\section{Evaluation}
\label{sec:eval}

The novelty of our approach and the problem we are tackling makes it hard to find suitable public data set or related work to compare with.
As mentioned in Section \ref{sec:historical-featurization} and Section \ref{subsec:context-types},
\fc relies on histories of access to uniquely identifiable resources in addition to the implicit social network structure of the company employees.
Unfortunately, to the best of our knowledge, no publicly available data set is available with such characteristics.



To perform an extensive evaluation of \fc, we tested it in the real-world \ggl environment.
This environment is orders of magnitude larger
in the number of users, resources, and access events than any
public data set; with more than $100,000$ full-time employees,
additional temps, vendors, and contractors; and 10 billion yearly events.
Most importantly, it displays the full, messy, long-tailed behavior of a large enterprise.
We used a company internal program where volunteers spend two weeks performing an attack (under supervision).
Our system is evaluated on the task of retrieving these attackers' activity against the
background of all other activity in the company.

%
%

We asked volunteers to try to access sensitive information outside of their normal
area of work, mimicking real-world insider threat scenarios.
Participants were instructed to choose and emulate a profile from a playbook 
containing the following attacker goals.
In all cases, the attackers seek to discover and access sensitive information about a particular topic.
\newcommand{\youtube}{Media Sharing Platform\xspace}
\newcommand{\pixel}{Hardware Product\xspace}
\newcommand{\deepmind}{AI Research\xspace}

\begin{itemize}
\item{\textbf{\youtube}:}
    Attackers seek information about the inner workings of the platform.
    This information could include corporate financial data, earnings for individual creators,
    or ranking algorithms.
\item{\textbf{\pixel}:}
    Attackers seek information about the company's next generation device,
     involving obtaining planning documents, design documents, timelines, pictures,
     or schematics.
\item{\textbf{\deepmind}:}
    Attackers seek information
    on next generation AI research and development projects.
    This includes documentation pertaining to current and future research projects,
    unpublished papers, source code, and model weights.
\item{\textbf{Security Research}:}
    Attackers target internal security research. This includes undisclosed (zero-day) vulnerabilities,
    either within the company or in third-party systems, which
    they could potentially exploit or sell to nation-state adversaries;
    as well as internal security analysis tools and associated documentation.
\end{itemize}

\begin{table}[t]
    \caption{
        \small
        Number of Attack-Related Events
    }
    \label{tab:attack-event-counts}
    \centering
    \footnotesize
    \begin{tabular}{ccccc}
         Target $\backslash$ Type
         & Docs & Tables & HTTP/RPC  & Total \\
         \toprule
         \youtube
            & \pz21
            & \pz20
            & 127
            & 168 \\
         \deepmind
            & \pz60
            & \pz\pz5
            & 260
            & 325 \\
         \pixel
            & 123
            & \pz\pz8
            & 176
            & 307 \\
         Security Research
            & \pz54
            & \pz\pz3
            & 123
            & 180 \\
         \midrule
         Total 
            & 258
            & \pz36
            & 686
            & 980 \\
        \bottomrule
    \end{tabular}
\end{table}

We deliberately did not provide detailed attack plans nor specific resources to target.
This approach relies on the participants' creativity and curiosity,
while ensuring a minimum level of independence
between our system and its evaluation.

All fifteen participants were full time employees at \ggl with a demonstrated interest in cybersecurity.
Most of them were software engineers and
all were familiar with internal tools and information storage practices.
Each scenario involved up to four participants,
and collaboration between attackers was both allowed and encouraged.
As we are simulating insider threats,
participants were not instructed to perform infiltration.
They also were instructed not to exfiltrate any data, as per company policy.
They simply began accessing sensitive data
using their preexisting credentials,
recorded their actions, and provided an after-action report.

To the best of our knowledge, an attack simulation and detection system evaluation with this level
of environmental and attacker realism, and scale, is unprecedented in the
published literature.
Using the attack report, investigators manually discovered and recorded
on the order of 1000 events in the \fc system that directly correspond to the attack.
Though they did not need to, some attackers performed some actions related to their usual responsibilities.
Actions such as these, and those that are only ambiguously related to the attack, are not counted here.
Table~\ref{tab:attack-event-counts} breaks the attack events down by target and action type. These are the same action types explained in Section \ref{sec:historical-featurization}.

\subsection{Single Event Detection Performance}
We train our baseline model using one year of data ending 90 days before the attack exercise begins.
We select hyper parameters using a spatially and temporally split
training and validation set, as per
Appendix
\ref{sec:spatial-split} and \ref{sec:hparam}.
The validation set is the two week period before the attack
exercise, i.e, there are 76 days of temporal separation.
After hyper parameters are selected, we train a model without
spatial splitting for evaluation.
Table~\ref{tab:dataset} summarizes key numerical characteristics of our training set.

\begin{table}[t]
\caption{Training set characteristics.}
\label{tab:dataset}
\centering
\footnotesize
\begin{tabular}{l  c} 
\toprule
 Covered time period & 1 year \\
 Distinct human principals & $\in [10^5, 10^7)$ \\
 Distinct resource identifiers & $\approx 10^8$ \\
 Actions before simplification & $\approx 10^{13}$ \\
 Actions after simplification & $\approx 10^{10}$ \\
 Compressed TFExamples & $\approx 3$ Tb \\
 \bottomrule
\end{tabular}
\end{table}

We use data gathered from the attack simulation
and evaluate a simplified model of our detection system: A team of security analysts who
audit a fixed proportion of events, those with the highest \fc scores.
We assume, for simplicity, that analysts will recognize attack events when they audit one.

The proportion of events audited is an upper bound on the false positive rate (FPR) of the system,
and the prevalence of attack events is small, so they are approximately equal.
Hence, we think of FPR and `requisite
audit proportion to discover an event'
as interchangeable for practical purposes.
We present receiver-operating characteristic (ROC) curves, which compare
FPR and recall, to
describe our retrieval of attack events.

While higher recall (ceteris paribus) is better, we focus on the low recall and low FPR region as
successfully detecting a single attack event will trigger an investigation that
can detect the rest of the events.
Additionally, many of the low-scoring attack events are relatively common and innocuous.
They include looking up potential victims who work in the target organization,
using \ggl's internal company directory.
This kind of reconnaissance activity,
without additional information from other events,
is difficult to identify as attack related for both analysts and \fc alike.

\begin{figure}[h]
    \centering
    \includegraphics[width=1.0\columnwidth]{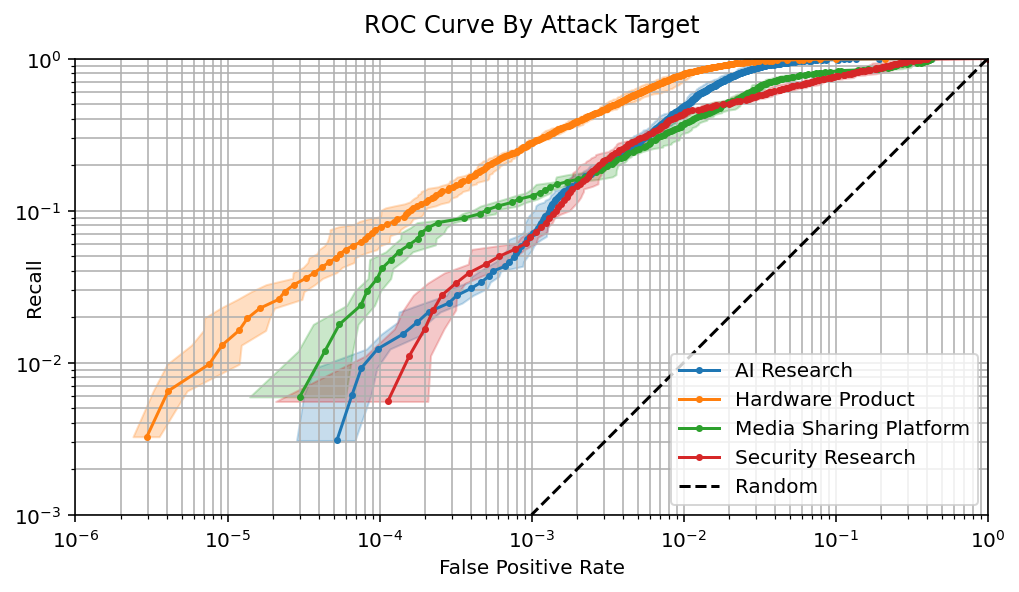}
    \caption{
        \small
        Simulated Attack ROC curves, broken down by attack target.
        The baseline model is trained 5 times. We take the 5 ROC curves
        and for each TPR, plot the mean associated FPR and shade the region
        between the minimum and maximum FPRs.
    }
    \label{fig:roc_target}
\end{figure}

Figure~\ref{fig:roc_target} demonstrates that
analysts can detect four attack events,
all targeting \pixel, 
while only auditing 1 in $10^5$ events,
as there are 4 points below $10^{-5}$ FPR.
If 1 in $10^4$ events are audited,
we detect almost 10\% of the \pixel attack events
and detect three of the four attack targets.
The highest ranked attack action
has an FPR below 0.0003\%.
That is detectable with an extremely low, 3 in $10^6$, audit proportion.

\subsection{Common Event Filtering}
\begin{figure}[h]
    \centering
    \includegraphics[width=1.0\columnwidth]{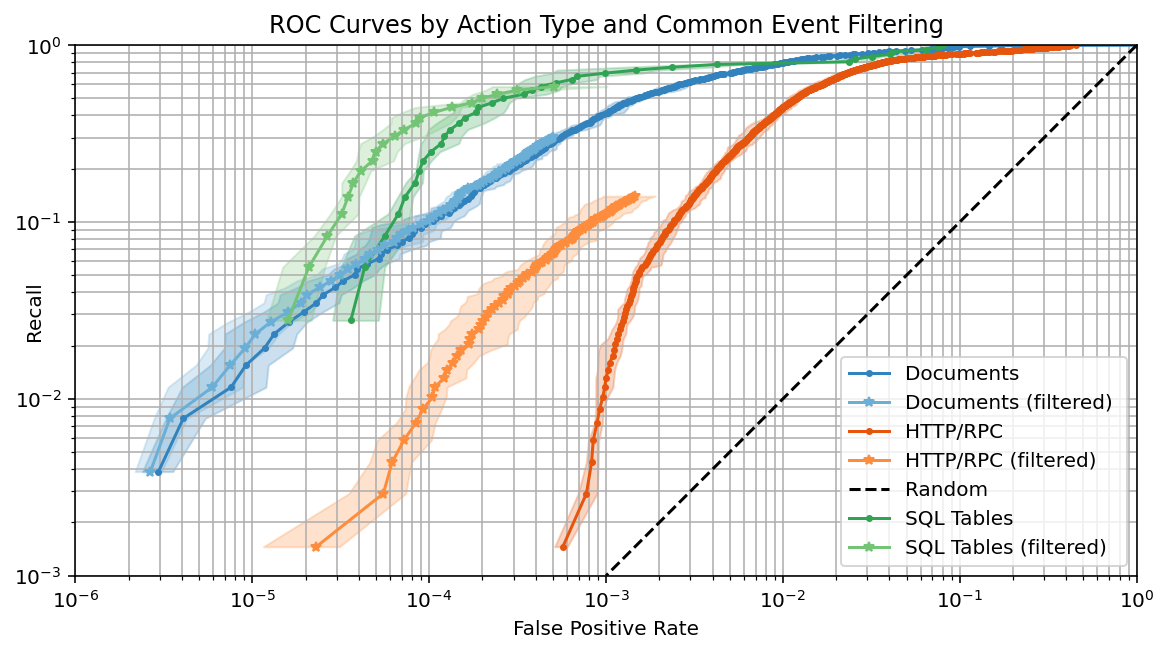}
    \caption{
        \small
        Simulated Attack ROC curves, broken down by action type
        and whether common event filtering is applied.
    }
    \label{fig:roc_action_type}
\end{figure}

Figure~\ref{fig:roc_action_type} illustrates
the effect of common event filtering (introduced in Section \ref{sec:common-filtering}) on single event detection varies by action type.
Attack related document access events are already discoverable
with $f$ alone using a 1 in $10^5$ audit proportion.
Consequently, common event filtering yields no noticeable enhancement for this particular action type.
Retrieval of attack-related SQL table access events is improved slightly, lowering
the requisite audit proportion for detection by almost a factor of 2.
Common event filtering most significantly affects retrieving
attack-related HTTP/RPC requests,
reducing FPRs at the low end by a factor over 20x.
It is required to retrieve such events
with a 1 in $10^4$ proportion used by the other types.

\subsection{Attacker Detection Performance}
Given the scale of $10^{10}$ events per year outlined in Table~\ref{tab:dataset}, security analysts would need to audit thousands of events per day to reach the 1 in $10^4$ audit proportion required to detect most targets of the simulated attack.

Additionally, because the top ranked events are generally unrelated,
analysts are fatigued by frequent context switching.
Overall, we find the continuous auditing of single events, by itself, prohibitively expensive for use in production.
In practice, the audit volume may be reduced with additional rules and signals which are not considered here.

Hence, we evaluate another simplified model of our detection system that
ranks \emph{principals} every day.
Security analysts audit the $K$ highest ranked principals who were not audited in the preceding 7 days.
For simplicity, we assume attackers are detected if they're audited.
We rank a principal, $p$, using $g(c_p, A_p)$ where $A_p$ is the set of all actions
performed by $p$ in the preceding 7 days, after common event filtering.

For comparison, we relax our idempotence to behaviorally redundant actions requirement (Section~\ref{sec:idempotence})
and consider simpler aggregation methods of the form:
$
h_{s}(A, c) = s(\{ f(a, c): a\in A\})
$
where $s\in\{\max,$ average, sum$\}$,
applied after post filtering.
Notably, these aggregations do not leverage behavioral embeddings
and averaging does not meet our monotonicity requirement either.

\begin{figure}[t]
    \centering
    \includegraphics[width=1.0\columnwidth]{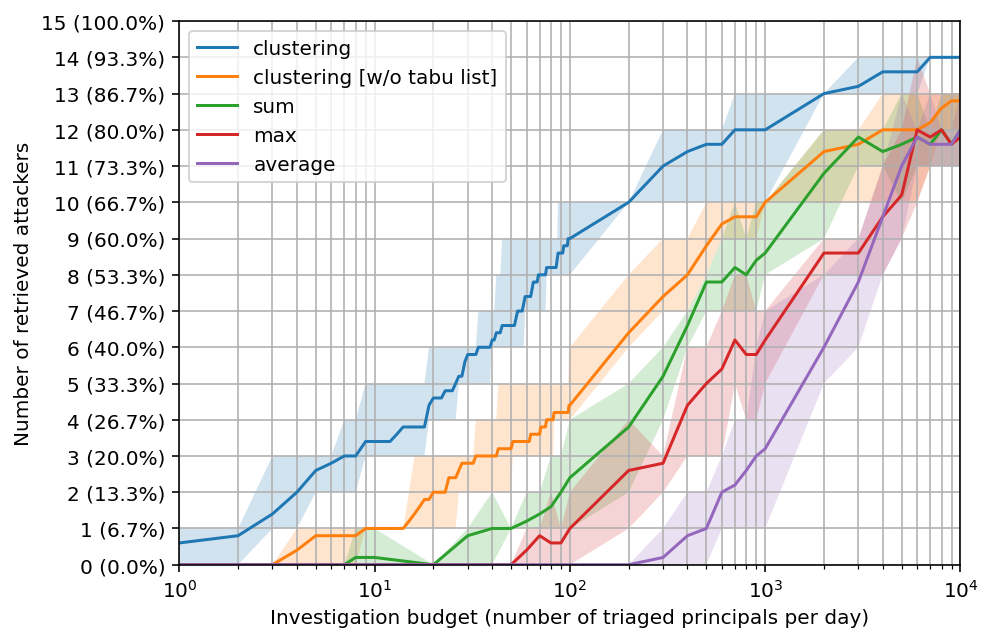}
    \caption{
        \small
        Number of retrieved attackers as a function of the daily investigation budget. For each aggregation strategy, a solid line denotes the mean over five distinct models; the minimum-maximum band is also shown.
    }
    \label{fig:aggregation_results}
\end{figure}

Figure~\ref{fig:aggregation_results} demonstrates that our clustering-based aggregator can detect 3 to 4 of the 15 attackers
at a cost of 10 audits per day.
This is remarkably low for a company with over $100,000$
full time employees and additional temporary workers, vendors, and contractors.
The alternative aggregation methods
require orders of magnitude larger auditing budgets to achieve similar results.
They are even sometimes non-monotonic as attackers are spuriously audited
during a period without attack activity (which does not count as detection)
and are prevented from being correctly audited for the subsequent week.

Preventing recently audited principals from being re-audited, \`a la Tabu Search~\cite{glover1990}, is usually helpful.
Principals' ranks exhibit strong auto-correlation across days because we apply
aggregated scoring over overlapping windows of activity. Preventing re-auditing uses
security analyst time more efficiently. We see in Figure~\ref{fig:aggregation_results} that
ablating the tabu list meaningfully reduces attacker recall at every investigation budget size.
Though, it still outperforms the naive aggregation techniques with tabu lists.

\subsection{Experiments}
\label{sub:experiments}
\begin{figure}[!ht]
    \centering
    \includegraphics[width=1.0\columnwidth]{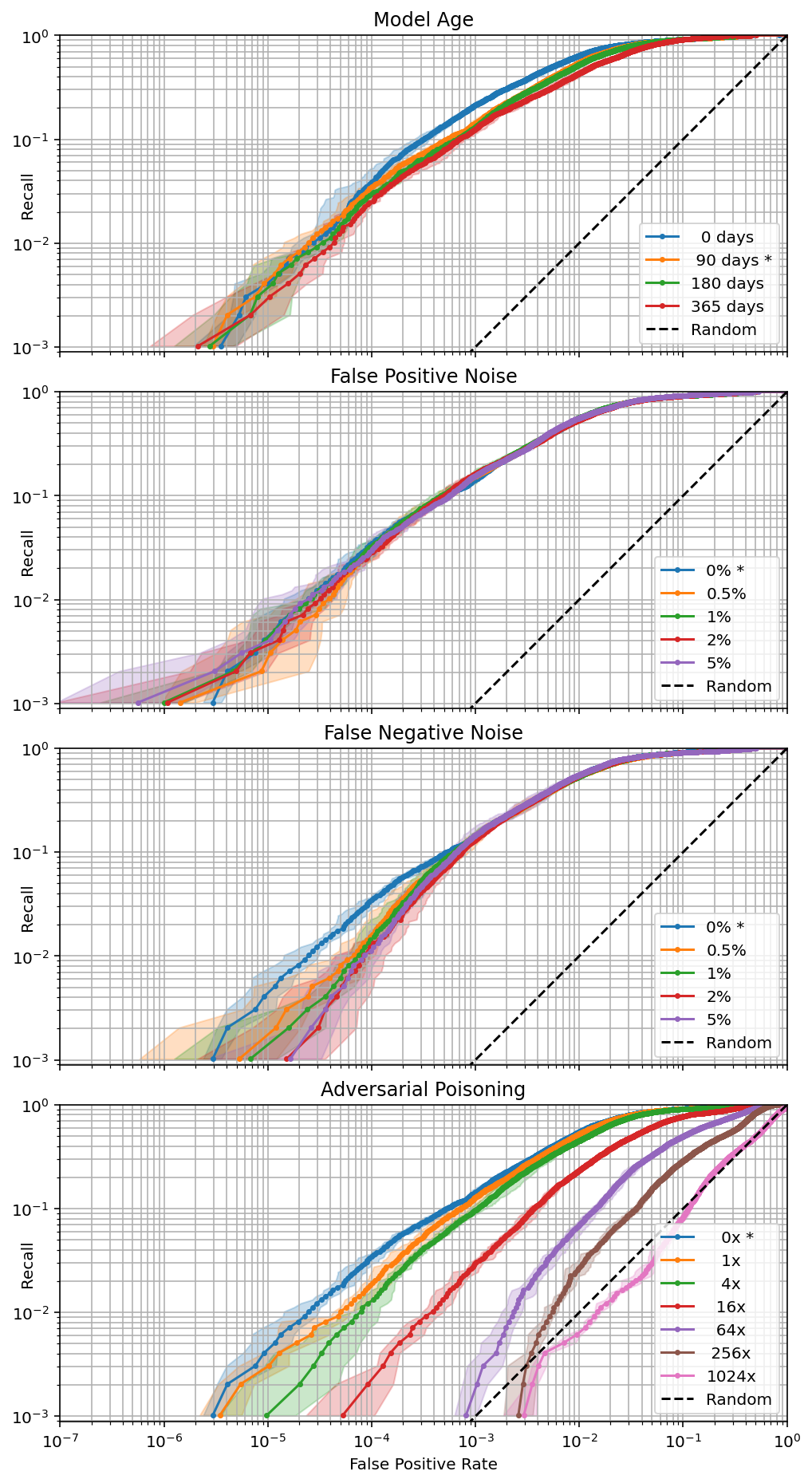}
    \caption{
        \small
        Simulated Attack ROC curves, for each experiment and experiment variation.
        Each model variant is trained 5 times. We plot the mean ROC curve and
        shade the region between the minimum and maximum.
        The variations that correspond to the baseline model are marked with *.
    }
    \label{fig:roc_all}
\end{figure}

Finally, we conduct experiments to evaluate our various claims about $f$
and show the results in Figure~\ref{fig:roc_all}.

\noindent \textbf{Resilience to Distribution Shift.}
We study the model's performance in the inductive setting and
the effect of distribution shift in our model age experiment.
We vary the length of the cooling-off period between
when the model's training data ends and when the simulated attack begins.
We train additional models whose cooling-off periods are 0, 180, and 365 days.
Figure~\ref{fig:roc_all} demonstrates that,
while newer models tend to perform slightly better in
the FPR region between $10^{-4}$ and $10^{-3}$,
there is similar performance in the FPR region below $10^{-4}$, where detection happens.
It is remarkable that even in an environment as dynamic as \ggl
with many team reorganizations, role transfers, new and cancelled projects,
we can run a 1 year old model without significantly degraded
detection performance.

\noindent \textbf{Spurious Negatives in Synthetic Positives.}
We evaluate the claim  that our model can tolerate spuriously generated natural pairs during positive sampling
in our false positive noise experiment (Section~\ref{sec:robustness-claim}).
We intentionally convert random negative pairs into positive pairs until $\phi\%$ of the final positive pairs are drawn from the negatives.
%
Figure~\ref{fig:roc_all} demonstrates that false positive noise up to $\phi=5\%$, not only does not diminish our ability to retrieve attack events, but seems to improve it.
The mean FPR of the highest scoring attack
event is below $10^{-6}$.
We believe the improvement is
related to label smoothing~\cite{szegedy2015-inception-labelsmoothing}, which regularizes the model's overconfidence in the negative events.

Next, we study the claim that \fc can tolerate some non-business-justified activity in the natural events (Section~\ref{sec:robustness-claim}).

\noindent \textbf{Robustness to Random False Negatives.}
In our false negative noise experiment, we study the effect of \emph{random} false negatives by intentionally converting random positive pairs into negative pairs until $\theta\%$ of the final negatives come from synthetic positives.
Figure~\ref{fig:roc_all} shows \fc can detect some attack
events while auditing just 1 in $10^{5}$ events when
$\theta < 1\%$.
However, performance clearly decreases as $\theta$ increases.
$\theta \geq 1\%$  is roughly equivalent to an environment in which
at least $1\%$ of the logs incorrectly attribute events to random principals,
which is generally considered unacceptable under normal operations.

\noindent \textbf{Robustness to Adversarial Poisoning.} In our adversarial poisoning data experiment, we anachronistically replicate the events from the simulated attack
into our training data set and mislabel them as natural.
This studies the worst-case effects of access abuse present in the historic access logs.
Figure~\ref{fig:roc_all} shows that we can still detect the attack simulation
with a 1 in $10^5$ audit proportion
if the attack events are replicated into in the training data once
or even up to four times.
At a 1 in $10^4$ proportion, attack events may be replicated up to 16 times.
However, retrieval is worse than random if the attack events are
replicated 256x or more.

We conclude that our assumption that most events
in the historical data are benign
is critical to \fc's retrieval capabilities.
Some false negatives, random or malicious, may be present.
So long as the proportion is sufficiently small, $f$ can still be used to retrieve attack events.

\subsection{Evaluation Summary} By using only historical data of normal business activity,
contrasted against sampled synthetic positives,
and no ground truth attack data,
we can train a model capable of retrieving attack events
in a real-world, large company environment.

\fc retrieves both events and principals
associated with a simulated insider attack
with an extremely low false positive rate, below $0.01\%$.
This performance is maintained even 1 year after model training and
when we intentionally violate our benign historical data assumption
with small quantities of random noise or poison data.
The false positive noise in our synthetic positive sampling
is not only tolerable, but seems to even improve performance.
The mean tail FPR for $\phi=5\%$ is less than $10^{-6}$.
This suggests we should research incorporating
label smoothing
\cite{szegedy2015-inception-labelsmoothing}
and weak supervised learning
\cite{Ratner_2017_weak_SL}.
In addition to improving baseline performance, we believe it may
let us relax our benign data assumption.

We evaluated two uses of our behavioral embeddings.
First,
common event filtering over single events can dramatically lower false positive rates, as with HTTP/RPC actions which have relatively weak baseline performance.
Second, we make our aggregator idempotent to behaviorally redundant actions,
which enables the detection of
\emph{attackers} at extremely small auditing budgets,
below $0.01\%$ of our workforce per day.


\section{Limitations}

\noindent \textbf{Recall vs. False Positive Rate.} There is a trade-off between recall and false positive rate, as shown in our evaluation using ROC curves. Increasing recall at the expense of a higher false positive rate could increase the number of detected attacks, but at the cost of requiring more security analysts and financial resources. Ultimately, the decision depends on an organization's budget and its risk tolerance. 

\noindent \textbf{Benchmark.} The sensitive nature of the data makes it impossible to share with the public. Furthermore, there are no suitable public benchmarks available. 
Consequently, we are unable to either share our data or utilize a benchmark for comparison. This is a limitation of a realistic evaluation in a production corporate environment rather than a constraint specific to this research.

\noindent \textbf{Prior Work.} Unfortunately, we do not have a point of comparison for our system. However, we believe this work will be a valuable contribution to the field of enterprise insider threat detection. To this end, we open sourced \fc's model code, along with a reference implementation of our action and context featurization pipelines. This enables implementations of \fc outside of \ggl and establishes a baseline for future insider detection systems.

\noindent \textbf{Contextual Detection Gap.} Because events are contextualized per user, \fc's main limitation is that it is not able to detect malicious actions that are indistinguishable from everyday user actions. This makes it unsuitable for detecting a rogue employee's access to sensitive files that are directly related and relevant to their work. \ggl relies on other defense mechanisms to address such threats.

\noindent \textbf{Generalizability of History Featurization.} History featurization requires a unique key to identify actions on the same resource over time. In practice, we find \fc to be more successful for documents and SQL tables than for HTTP and RPC services. This may be because hostnames (and specialized URL parsing) are suboptimal choices for history key. Devising a better key system for those and for security-relevant actions which do not have obvious keys, such as ad-hoc CLI invocations, remains an open question.

\noindent \textbf{Non-human identities.} \fc primarily featurizes principals by their peers, as identified by meetings, code review patterns, and the HR reporting chain. This assumes that they are humans with coworkers and does not 
apply to non-human identities such as service accounts or AI agents. While we believe our technique, using contrastive learning with positive sampling, may be adapted to the non-human domain, effective featurization of such identities also remains an open question.

\noindent \textbf{Adversarial Robustness.}
To decrease a single-event score, attackers may use training or testing time strategies.
In the simplest training-time strategy, the attacker may repeatedly perform an action in the hopes of polluting the training set and in turn poisoning the model.
Section~\ref{sub:experiments} quantifies \fc’s robustness to this type of attack: the system tolerates some adversarial noise before eventually degrading.
In testing-time attacks, the adversary manipulates their context and action features so as to decrease the score of the event.
Our particular choice of context features make it possible for an attacker to partially manipulate their content, for instance by scheduling meetings or performing code reviews with specific, targeted peers.
However, we posit that the publicly-visible and social nature of such actions create a barrier for the attacker: creating meetings with or performing code reviews for highly-unusual peers might raise questions and trigger additional scrutiny for the attacker.
Finally, on the action side, an actor repeatedly accessing the same resource will decrease the anomaly score, as illustrated in Figure~\ref{fig:history}, though the first access still scores highly.
As a last line of defense, the \fc system is one among many other internal detection mechanisms: attempts to evade detection may make the attacker paradoxically more detectable by more regular, rule-based systems.

\section{Related Work}

We now discuss the prior work that contributed similar techniques and capabilities in the field of anomaly detection applied to access abuse.





\noindent \textbf{Single event detectors.}
Gafny \etal address the problem
of anomalous resource access by learning invariant 
relationships between resource and principal features that are verified
for every access~\cite{gafny2011}. 
This uses frequent itemset mining on
categorical features~\cite{agrawal1993}.

Gelven \etal consider the problem of anomalous
principal-resource access detection, where both principals and resources
are opaque nodes of a bipartite graph~\cite{gelven2023}.
The system uses matrix factorization on historical usage data
to perform link prediction, in a classic collaborative filtering fashion.
\fc draws from two-tower deep recommendation systems ~\cite{covington2016}, allowing for rich featurizations of both principals and resources.

Zheng \etal also observe the structural similarity between
anomaly detection and recommendation systems~\cite{shadewatcher}. They apply graph neural network
over provenance graphs and retrieve low probability interactions between system entities.
\fc operates at a higher level of abstraction, reasoning about authenticated users
and distributed resources in a corporate environment, rather than processes and files.
Another difference is that \fc uses features entities with weighted sets of tokens,
as opposed to per-entity embedding vectors,
so it can better featurize out-of-distribution entities.

\noindent \textbf{Detection of Anomalous Batches of Events.}
Many anomaly detection systems operate solely on batches of activity
(e.g., login sessions, user-days, user-weeks) and rely on
aggregate features, typically counts of events.
These numerical features are in turn fed to traditional unsupervised 
learning techniques.
Good examples include principal component analysis on network traffic
patterns\cite{huang2007} and auto-encoders on network
flows\cite{kathareios2017}.
Closer to insider threat detection, several studies~\cite{salem2011,senator2013} train classical statistical anomaly detection models, such as one-class SVM~\cite{manevitz2001}, on a per-user activity basis,
using aggregated event counts and other volumetric behavioral features.
The system described by Liu \etal detects anomalous user-day batches,
and also produces per-event anomaly scores~\cite{liu2019a}.

Contemporary work uses deep-learning models on either manually-defined
count-like features describing the overall
activity session~\cite{liu2018,gayathri2020,le2021} or directly on
the sequence of
actions types~\cite{tuor2017,yuan2019,macak2020,zhang2021}, without
considering the specific resources accessed.

Unlike the above, \fc can evaluate the anomalousness of \emph{individual} events, and represents resources by their
access history. It also can aggregate sets of events to achieve low FPR detection over batches of activity.

\noindent \textbf{Detection by Aggregating Single Events.} Hassan \etal designed an algorithm 
applied to arbitrary detection and alerting systems
that, given a provenance graph, aggregates alerts using that graph~\cite{nodoze}.
Asaheel \etal applied sequence models atop provenance graphs
to extract an `attack story' from the larger graph~\cite{atlas}.
It requires labeled data of previous attacks.
\fc does not require an explicit model of 
relationships between events.
It can perform detection over both individual events and
sets of events associated with a single principal.

\noindent \textbf{Feature Representation.} Gates \etal design an anomalous file access detector~\cite{gates2014}.
This work is file system specific, though like \fc, they use users' access history as features.

Some systems maintain per-user,
variable size non-ML profiles, and compare the daily activities
against those using ad-hoc hierarchical or graphical data representations~\cite{parveen2011,legg2017,toffalini2018}.
Contemporary work uses deep-learning models on either manually-defined
count-like features describing the overall
activity session~\cite{liu2018,gayathri2020,le2021} or directly on
the sequence of
actions types~\cite{tuor2017,yuan2019,macak2020,zhang2021}, without
considering the specific resources accessed.
Scores from Liu \etal's system\cite{liu2019b}
are derived from the word2vec-learned~\cite{mikolov2013}
representations of the log entries and are not evaluated for
separate uses.

Mathew \etal adopt a learned, per-role profile
context for detecting improper data access at the database level~\cite{mathew2010}.
Their features
are derived from the SQL query response itself, which we consider to be
content-based, as opposed to our history based featurization.

\section{Conclusion}
\label{sec:conclusion}
We have presented \fc, a deep contextual anomaly detection system that helps protect \ggl against malicious or unintentional insider threats with an extremely low false positive rate. Three techniques enable \fc to surface fundamentally novel insider threats with very high precision at scale.

First, positive sampling forgoes the need for expensive attack data collection for training purposes.
Instead, training exclusively relies on cheap and readily available historical activity logs.
Second, \fc works across resource types (arbitrary cloud documents, data-stores as they appear in SQL-based queries, and HTTP/RPC requests to internal network resources) and generalizes to unseen-at-training-time users and resources.
This is achieved by using historical access-based and peers-based featurization strategies of actions and contexts, which in turn eliminates the need for frequent model retraining.
Third, \fc embeddings enable measuring the behavioral similarity of principals and resources, and unlock simple yet powerful filtering and aggregation methods.

To our knowledge, \fc is the first proposed contextual anomaly detection system for insider threats with sufficient precision to be deployed in a large-scale environment.

\section*{Acknowledgments}
We gratefully acknowledge the contributions of the following individuals, whose support, feedback, and dedication
were essential to this research: Heather Adkins, Aman Ali,
Nicholas Capalbo, Ceri Driskill, Marius Killinger, Louis Li,
Peng Liu, Justin Ma, Tom MacGregor, Eric Morley, Michael Sinno, Johan
Strumpfer, and Zachary Westrick.

\section*{Ethical Considerations}
\label{sec:responsible-ml}
We are committed to the responsible and fair usage of machine learning.
By detecting insider threats, the system participates in the
responsible handling of the end user data of billions and the
protection of the company's intellectual property.

Our commitment to employee privacy and fairness constrains what data we allow ourselves to use:
In Section~\ref{subsec:action-types}, only action types
that involve potentially sensitive activities on the corporate systems are instrumented.
In Section~\ref{subsec:context-types}, we derive context features
from internally-public information.
In particular, the collected data is limited to the corporate environment and excludes any personal accounts or general browsing activity, and we only use meeting data where the meeting is set to be visible to anyone internally: private or hidden meetings are excluded.

Additionally, we check for the presence of bias by empirically validating that
certain categories of employees are not over-represented in the final detection results.
We make our methodology public inside the corporate environment.
Model weights and feature values are kept internally private.

\section*{Open Science}
\label{sec:open-science}

We fully support the goals of open science and the principles of reproducibility and shared community advancement.
To this end, we are committed to open sourcing the parts of \fc that we can, including the code that
defines, trains, and evaluates our model.
Therefore, we open source the core \fc model code and the code for constructing input features\footnote{
\begin{tabular}{l}
\url{https://github.com/google/facade}\\
\url{https://zenodo.org/records/20316179} (historical record) 
\end{tabular}
}. The open sourced version is implemented in the Python language as a reference implementation.

This paper also provides comprehensive descriptions of our techniques, model architecture, and feature engineering.
We believe these artifacts will be valuable to the research community, enable others to implement similar systems, and facilitate future innovations.

However, \ggl's business considerations are such that
we are unable to share the version of the \fc model
that is evaluated in this paper, which is trained on \ggl's internal corporate data.
The same considerations apply to the underlying training and evaluation data, as well as the data collection pipelines themselves.
These artifacts
cannot be directly used to secure systems outside of \ggl and
contain sensitive information that could be exploited for purposes that do not align with our business.

For example, while the number of full time employees is public information,
\ggl does not disclose the number of temporary employees, vendors, or contractors that it hires.
This makes the exact number of principals confidential information.
As a result, in Table~\ref{tab:dataset}, we reveal the number as a two order of magnitude range.
%
We featurize both actions and contexts with weighted sets of usernames. Releasing the trained model reveals this confidential information. 
Business restrictions aside, these usernames are internal to our company,
limiting the public value of releasing the trained model.

\fc is trained on historic access logs.
We cannot release this training data because such logs reveal sensitive internal information; including the number of documents we produce, their rate of production, and their rate of access.
There are similar concerns for SQL tables and internal URL access data too.
If the principals are deanonymized, access data reveals \ggl's relative investment and staffing into different business areas,
which is considered highly confidential information.

Finally, our data collection pipelines are tightly coupled with \ggl's internal logging infrastructure.
Sharing the code provides little value to the scientific community as it will be unusable outside of \ggl.
Regardless, data pipelines are not what makes our contributions novel.


\bibliographystyle{plainurl}
\bibliography{bibliography}

\appendix

\section{Proof of Theorem 1}
\label{sub:proof}
\begin{proof}
As $\bA$ and $\bC$ are assumed to be finite,
the set of all possible models, $\mathcal{M}$, is $\bR^{n_a\times n_c}$.
Each model is parameterized by
$n_a n_c$ real weights $\{w_{i,j}\}_{i\in\mathbb{A},j\in\mathbb{C}}$. Any
classifier $f\in\mathcal{M}$ is of the form $f(a,c) = w_{a,c}$.

For generality, let $\alpha > 0$ be some weight assigned to synthetic positive instances.
The risk of a given classifier $f\in\mathcal{M}$ is defined as:
\[
R(f) = \sum_{a \in \mathbb{A}, c \in \mathbb{C}} P_{-}(a, c) \ell(-f(a, c)) + \alpha P_{+}(a, c) \ell(f(a, c))
\]
where $P_-$, $P_+$ respectively denote the probabilities of observing the pair as a
negative and as a synthetic positive.
By definition, we have $P_- = P$.

For $P_+$, notice that in the simplified version of positive sampling,
the synthetic positive pairs are formed by
randomly and independently drawing two natural action-context pairs $(a, c)$
and $(a',c')$ from the distribution $P$, resulting in the positive instance
$(a, c')$.
Hence, the positive distribution is the product of the marginal
action and context distributions $P_+(a, c) = P(a)P(c)$.

Plugging in the model form, we get:
\[
R(f) = \sum_{a \in \mathbb{A}, c \in \mathbb{C}} P(a, c) \ell(-w_{a,c}) + \alpha P(a)P(c) \ell(w_{a,c})
\]
Because of the simplicity of the model form, finding the model weights that minimize the
risk is a separable optimization problem, by action-context pair.
For a given pair $(a, c)$, an optimal weight $w^*_{a, c}$ minimizes:
\[
x \mapsto P(a, c) \ell(-x) + \alpha P(a)P(c) \ell(x)
\]
Moreover, by convexity of the objective function, there exists a unique such weight,
and the derivative of the above is such that:
\[
-P(a, c) \ell'(-w^*_{a,c}) + \alpha P(a)P(c) \ell'(w^*_{a, c}) = 0
\]
Rearranging the terms, we get:
\[
\frac{\alpha \ell'(w^*_{a, c})}{\ell'(-w^*_{a,c})} = \frac{P(a, c)}{P(a)P(c)}
\]
Finally, notice that the function
\[
x \mapsto \frac{\alpha \ell'(x)}{\ell'(-x)}
\]
is a ratio of negative strictly increasing and negative strictly decreasing functions,
making this a positive strictly decreasing function. Therefore, letting $f^*$ be the
classifier which minimizes the risk, we have that for any two pairs $(a, c)$ and $(a', c')$:
\begin{align*}
\frac{P(a, c)}{P(a)P(c)} > \frac{P(a', c')}{P(a')P(c')} &\Leftrightarrow \frac{\alpha \ell'(w^*_{a, c})}{\ell'(-w^*_{a,c})} > \frac{\alpha \ell'(w^*_{a', c'})}{\ell'(-w^*_{a', c'})} \\
&\Leftrightarrow w^*_{a,c} < w^*_{a', c'} \\
&\Leftrightarrow f^*(a, c) < f^*(a', c')
\end{align*}
\end{proof}

\section{Choice of Loss Function}
\label{subsec:lossfn}
Having reduced the previously unsupervised problem into a supervised learning task,
we turn our attention to the choice of loss function.
While the task of scoring pairs of actions and contexts according to their affinity
is drawn from contrastive learning~\cite{chopra2005}, this application
domain has specific issues to consider when designing an appropriate loss function.

We expect a small amount of label noise.
Although Theorem \ref{thm:lift} applies asymptotically,
in practice, our training data is finite and our models do not have infinite capacity.
The presence of labeling noise can, at best, delay training convergence
and, at worst, introduce model bias~\cite{chuang2020}.
Therefore, we avoid loss functions that place excessive emphasis on high-scoring negatives or low scoring positives.
We prefer loss functions with a bounded gradient,
ruling out the soft-nearest neighbors loss~\cite{salakhutdinov07,frosst19} and the
original contrastive loss~\cite{chopra2005}.
Ideally, the loss function's emphasis on high scoring negatives would be parameterized.
Label noise also restricts the applicability of hard instance mining~\cite{robinson2021}.

Our goal is to identify the most anomalous action-context pair overall,
without a priori fixing the action or the context.
Therefore, scores must be commensurable among all context and action pairs,
ruling out 
anchor-based loss functions like the triplet loss~\cite{schroff2015},
as such loss functions may result in anchor-dependent score scales.

A final factor that influences our loss function design is the underlying data imbalance,
with potentially orders of magnitude more synthetic positives than natural instances.
Pairwise ranking loss functions behave well in such scenarios
as they are, by definition, insensitive to such imbalance.

Taking into account the above requirements, we use the following pairwise loss function:
\[
\mathcal{L}(B) =
    \left(
    \frac{1}{N} \sum_{i=1}^N 
    \left(
    \frac{1}{P} \sum_{j=1}^P
    \ell\left(h + \frac{y^+_j - y^-_i}s\right)
    \right)^\omega
    \right)^{\frac1\omega}
\]
Where $B=(\{y^-_i\}_{i=1}^N, \{y^+_j\}_{j=1}^P)$ are the $N$ negative and $P$ positive scores in a training minibatch and $\ell$ is a smooth
pointwise loss function as in Theorem~\ref{thm:lift}. The parameter $\omega\in\bR^+$ controls the emphasis on high scoring negatives, and 
$s\in\mathbb R^+$ and $h\in\mathbb R^+$ are soft and hard margin parameters.
We use the following Huber-like \cite{Huber64} expression for $\ell$,
\[
\ell(t) = \begin{cases}
-t - \frac{1}{2} &\text{if } t < -1 \\
\frac{1}{2} t^2 &\text{if } t \in [-1; 0] \\
0 &\text{if } t > 0
\end{cases}
\]

When $\omega=1$, Theorem 2 in~\cite{gao2015} shows that
the loss is consistent with AUC minimization given our conditions on $\ell$.
In practice, we set $\omega$ using hyper parameter tuning, as per Appendix~\ref{sec:hparam}.


\section{Shortcut Learning}
\label{subsec:shortcut}
Shortcut learning is a form of overfitting to our contrastive learning objective.
It happens when the model learns to distinguish the natural pairs
from synthetic pairs using information that is ultimately irrelevant to the security
domain.

In Appendix \ref{subsec:shortcut-example}, we demonstrate a pathological featurization
and model that
by exploits the fact that the action and context principal are not
equal for synthetic positive pairs.
It can perfectly identify whether a given pair is natural or synthetic,
but does not achieve the end goal of detecting malicious activity.

Rich featurization schemes increase the risk of inadvertent shortcuts.
For instance, consider the action space of HTTP requests on internal websites,
featurized with a bag-of-words scheme on the requests' parameter values
and header data.
It is extremely likely that these tokens contain some high-entropy, principal-identifying
information that is irrelevant for behavioral analysis, for instance user-agent strings,
authentication cookies, HTTP query parameters that directly identify the principal.

Shortcut features are not restricted to those that identify the principal.
Shared temporal information between actions and contexts can also result in shortcut learning.
For instance, if the action featurization contains the time of the event,
and the context featurization contains the time at which the context was computed,
and provided context extraction is frequent enough, then positive sampling will
focus the model on matching the action and context timestamps.

We take care to ensure that undesirable feature shortcuts between actions
and contexts are not present.
In Appendix~\ref{sec:spatial-split}, we discuss an evaluation method that
detect some instances of shortcut learning.

\section{Shortcut Learning Example}
\label{subsec:shortcut-example}
Unfortunately, we demonstrate that it is possible to achieve perfect separation with positive sampling,
but not achieve the end goal of detecting malicious activity.
Consider the following pathological model, $f_s$ and featurization:
Both actions and contexts are directly and exclusively featurized by the token
of their principal, so that the labeled data set is:
\begin{align*}
\tilde{\mathcal{D}}_{\text{shortcut}} = &\{(p(a_i),p(c_i),-1)\}_i \\
\cup &\{(p(a_i),p(c_j),+1)\}_{i,j}
\end{align*}
where $p(x)$ denotes the username token of the principal associated with the action
or context argument $x$.

Let $\mathit{Users}$ be the set of distinct principal tokens $\{p_i | 0 \leq i < n\}$
and let $d\geq 2$ be the number of embedding dimensions.
The model consists of a non-negative embedding matrix $w \in \mathbb{R}_+^{M \times d}$,
where $M >> |\mathit{Users}|$.
$f_s$ hashes the principal token feature to select a row of $w$
to represent the context and also for the action.
It outputs their cosine distance, $d_{cos}$.
Let $h$ be a fixed pseudo-random function that maps principal tokens
to $[0; M-1]$. The complete model is thus:
\[
f(a, c) = d_{cos}(w_{h(p(a))}, w_{h(p(c)))}) 
\]

Even before any learning takes place, this model is already almost perfect from the
standpoint of positive sampling. Provided any reasonable random initialization scheme
for $w$,
\begin{enumerate}
    \item any natural pair $(a, c)$ is such that $f(a, c) = 0$, because both the action
    and context map to the same principal token, same embedding index and therefore
    same embedding, and
    \item any synthetic positive pair $(a', c')$ is such that
    $f(a', c') > 0$ with probability $1 - \frac{1}{M}$. This is because synthetic positives are
    by definition such that $p(a') \neq p(c')$, therefore the principal tokens map to
    different indexes with the above hash collision probability.
    The specific expected distance depends on the initialization scheme but
    increases with $d$.
\end{enumerate}

While $f$ almost perfectly separates natural from synthetic action-context pairs,
it is also evident that $f$ does not provide any information on whether a given document
access is anomalous for a user. It uniformly scores 0 any actual access.
In this scenario, we say that positive sampling is misaligned with the security
application domain.

\section{Spatial-Temporal Splitting}
\label{sec:spatial-split}
In our experience, \emph{spatially} and \emph{temporally} splitting the training
and validation sets is effective for revealing
when a model has shortcut learning issues, as
described in Section~\ref{subsec:shortcut}.

\begin{figure}[t]
\centering
  \includegraphics[width=1 \columnwidth]{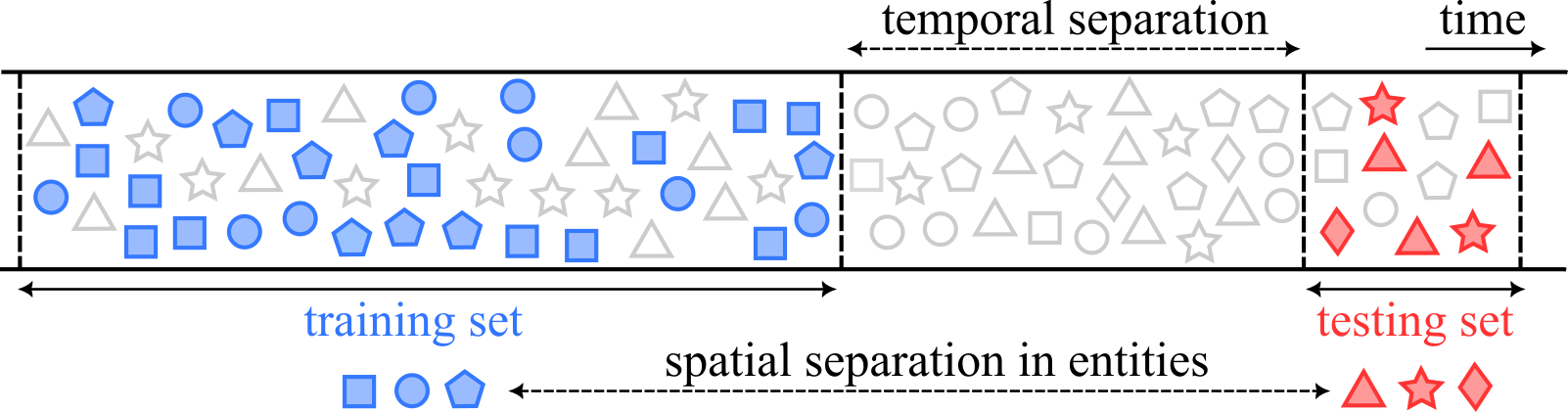}
\caption{
    \small
    Validation setup for tuning model architecture and hyperparameters.
    To properly assess model generalization performance in the presence of temporal
    distribution shift and of entities unseen at training time, we separate the training
    and testing sets in both time and ``space'' dimensions.
}
\label{fig:eval_setup}
\end{figure}

Our validation set is constructed to be disjoint from the training set as follows:
Temporally, the last event of training precedes the first event of testing set.
Spatially, we randomly hold out 50\% of the entities from the training set.
This is achieved by dropping all events pertaining to either to held-out principal
or to a held-out resource, using a deterministic pseudo random selection function
applied to their identifier string.
Usernames that were spatially held out are not present in the histories nor implicit social networks
of actions and contexts in the training set.
For the validation set, we only consider events where the user or the accessed resource was
held-out from the training set.
These ``new'' users and resources are featurized by the usernames that were \emph{not} held out of the training set.
Figure~\ref{fig:eval_setup} summarizes the approach.

\section{Hyper Parameter Optimization}
\label{sec:hparam}

We make extensive use of Vizier~\cite{golovin17}
to select the various hyper-parameters present across our system.
Vizier selects hyper parameters for $f$
to maximize the AUC on a spatially and temporally split validation set.

After $f$ is selected,
a separate Vizier study selects hyperparameters for common event filtering
and the aggregator, $g$. 
It is tasked with retrieving a set of 100 synthetic `attackers'.
We sample a random principal, and for each action type, 0 to 33 of their actions.
Those actions are interspersed amongst those of another random principal,
which we call the synthetic `attacker'.
Vizier optimizes the mean inverse log rank of synthetic attackers.
For computational efficiency,
we set $m$, the number of principles required to filter an event, to $1$
and have Vizier choose a minimum score threshold for
each action type. Only actions from events with a higher score are presented to the aggregator, $g$.

\end{document}